\documentclass[prb,twocolumn,secnumarabic,nobalancelastpage,amsmath,amssymb,
nofootinbib]{revtex4-1} %
\usepackage{mathptmx}
\usepackage{times}
\usepackage{graphics}      % standard graphics
\usepackage{graphicx}      % alternative graphics
\usepackage{url}           % for on-line citations
\usepackage{bm}            % special 'bold-math' package
\usepackage[hidelinks,colorlinks=true,
            linkcolor = cyan,
            urlcolor  = cyan,
            citecolor = cyan,
            anchorcolor = cyan]{hyperref}
\usepackage{amsmath}
\usepackage{subcaption}
\usepackage[font={small}]{caption}

\graphicspath{{pics/}}
\begin{document}
\title{\textit{Ab initio} study of electron mean free paths and thermoelectric properties of lead telluride}

\author{Qichen Song}
\author{Te-Huan Liu}
\author{Jiawei Zhou}
\author{Zhiwei Ding}
\author{Gang Chen}
\email {gchen2@mit.edu}
\affiliation{Department of Mechanical Engineering, Massachusetts Institute of Technology, Cambridge, Massachusetts 02139, USA}
%\date{\today}

\begin{abstract}
Last few years have witnessed significant enhancement of thermoelectric figure of merit of lead telluride (PbTe) via nanostructuring. Despite the experimental progress, current understanding of the electron transport in PbTe is based on either band structure calculation using first principles with constant relaxation time approximation or empirical models, both relying on adjustable parameters obtained by fitting experimental data. Here, we report parameter-free first-principles calculation of electron and phonon transport properties of PbTe, including mode-by-mode electron-phonon scattering analysis, leading to detailed information on electron mean free paths and the contributions of electrons and phonons with different mean free paths to thermoelectric transport properties in PbTe. Such information will help to rationalize the use and optimization of nanosctructures to achieve high thermoelectric figure of merit.
\end{abstract}

\maketitle

%%%%%%%%%%%%%%%%%%%%%%%%%%%%%%%%%%%%%%%%%%%%%%%%%%%%%%%%%%%%%%%%%%%%%%%%%%%%%
\section{Introduction}
The thermoelectric devices directly convert heat into electricity and can be used in the power generation application. The efficiency of the thermoelectric devices is determined by the material's figure of merit $zT =\sigma S^2 T/\kappa$, where $\sigma$ is the electrical conductivity, $S$ is the Seebeck coefficient, $\kappa$ is the thermal conductivity consisting the contribution from electrons ($\kappa_e$), ambipolar diffusion ($\kappa_{bp}$) and phonons ($\kappa_{ph}$), and $T$ is the temperature.
Several groups reported high figure of merit in PbTe through different nanostructuring approaches \cite{sidopepbte}\cite{Biswas2012}\cite{C0EE00456A}\cite{C2EE21536E}\cite{ZHANG2016572}. One beneficial feature of PbTe is its low intrinsic thermal conductivity due to the strong anharmonicity\cite{Delaire2011}\cite{Lee2014}\cite{PhysRevB.85.184303}. In addition, PbTe has low effective mass and multiple valleys, which give rise to its high electrical conductivity\cite{C2EE21536E}\cite{martinez1975electronic}. Even though a peak $zT$ as high as 2 has been achieved in $p$-type PbTe\cite{Hsu818}, further improvement of $zT$, especially in $n$-type PbTe, is desirable for thermoelectric energy conversion to be competitive\cite{C1EE02497C}\cite{Vining2009}\cite{tritt_chen_2008}.

To make nanostructure approach more efficient\cite{PhysRevB.53.R10493}, it is generally believed that one should design the nanostructures such that the major heat carriers, phonons, are strongly scattered by interfaces while the charge carriers, electron/holes, are barely affected\cite{Biswas2012}. The success of the nanostructuring technique is partially attributed to the disparity between the mean free path of electrons and phonons. In silicon, for example, the electron mean free paths are around tens of nanometers, while phonons have mean free paths up to a few microns. As a result, nanostructures with grain sizes between the electron and phonon mean free path strongly scatter phonons and reduce thermal conductivity dramatically yet have minimal effects on the electrical transport\cite{qiu2015first}. For PbTe, the thermal transport has also been examined from the first principles yielding that phonons with mean free paths smaller than 10 nm contribute majority of the thermal conductivity\cite{PhysRevB.85.184303}. However, its electron transport properties and electron mean free paths are much less understood. Past work have mostly employed the constant relaxation time approximation when studying the electrical transport properties of PbTe\cite{PhysRevB.81.195217}\cite{PhysRevB.82.195102}.
Although good agreements with experiments have been achieved for the Seebeck coefficient, the detailed information on the charge carrier dynamics remains unknown. In particular, by adopting the single/double Kane band model together with multiple scattering mechanisms, past work successfully explain the trend of the experimental findings\cite{vineis2008carrier}\cite{doi:10.1063/1.4973292}\cite{Ravich1968}\cite{Ravich1971a}, yet the analysis requires the fitting parameters extracted from experimental results thus not necessarily unveiling the accurate physical pictures.

In this article, we evaluate the electron scattering rates and electron mean free paths due to electron-phonon interaction using first principles for $n$-type PbTe. Due to the large mismatch in energy between electrons and phonons, a very dense $k$-point mesh is needed in the search of possible electron-phonon scattering modes such that energy and momentum conservation can be satisfied. To calculate the electron-phonon coupling strengths on a very dense mesh with acceptable cost, we adopt the interpolation scheme based on electron Wannier functions\cite{PhysRevB.76.165108}\cite{PhysRevB.95.075206}\cite{doi:10.1021/acs.nanolett.5b05288}\cite{qiu2015first}. By further comparing the mean free paths of electrons with phonons, we are able to thoroughly examine the electron transport and phonon transport in PbTe at the same time. The detailed spectral information on the thermoelectric properties of PbTe not only provides microscopic pictures of the dynamics of electron and phonon but can be used to rationalize the design of the nanosturctured PbTe to decouple electron and phonon transport in order to boost the thermoelectric performance.

%%%%%%%%%%%%%%%%%%%%%%%%%%%%%%%%%%%%%%%%%%%%%%%%%%%%%%%%%%%%%%%%%%%%%%%%%%%%%

%\begin{figure*}[t]
%\includegraphics[width=.9\linewidth]{ephmat_combine_ph}
% \caption{(a) The electron-phonon coupling matrix element for electron in the lowest conduction band at L point interacting with phonons along the high symmetry lines of the first Brillouin zone. Solid lines are matrix elements considering the screening effect and dashed lines are matrix elements without considering the screening effect. The lines corresponding to the two different cases overlap with each other except for the long wavelength LO phonons (black dashed line and yellow solid line near the $\Gamma$ point ). The dopant concentration is $\mathrm{10^{18} cm^{-3}}$. (b) }
%\label{ephmat}
%\end{figure*}

\section{Methods}

\subsection{Electron transport properties}

The charge flux and the heat flux are correlated with the temperature gradient and electrochemical potential gradient by the transport coefficients\cite{chen2005nanoscale},
\begin{equation}
J_c =  -L_{11}\left(\frac{1}{q}\mathbf{\nabla}_\mathbf{r}\Phi\right)-L_{12}\mathbf{\nabla}_\mathbf{r}T.
\label{jc}
\end{equation}
\begin{equation}
J= -L_{21}\left(\frac{1}{q}\mathbf{\nabla}_\mathbf{r}\Phi\right)-L_{22}\mathbf{\nabla}_\mathbf{r}T.
\label{j}
\end{equation}
The first term in Eq.~\ref{jc} describes the electrical current due to the electrochemical potential gradient, and the coefficient $L_{11}$ is the electrical conductivity, which can be derived from the linearized Boltzmann transport equation for electron under the relaxation time approximation as,
\begin{equation}
\sigma_{\alpha \beta}=L_{11}=-\frac{q^2}{\Omega N_{n\mathbf{k}}}\sum_{n\mathbf{k}}\mathbf{v}_{n\mathbf{k}\alpha}\mathbf{v}_{n\mathbf{k}\beta}\tau_{n\mathbf{k}}\frac{\partial f_{n\mathbf{k},0}}{\partial \epsilon_{n\mathbf{k}}},
\end{equation}
where $\alpha$ and $\beta$ are certain directions in Cartesian coordinates and $\Omega$ is the volume of a unitcell. By changing the condition for the summation from \{$n\mathbf{k}$\} to \{$n\mathbf{k},|\mathbf{v}_{n\mathbf{k}}|\tau_{n\mathbf{k}}<\lambda$\}, we obtain the contribution to the conductivity of electrons with mean free paths up to a given value $\lambda$. Note that we can break the summation into the summation over electron states and hole states separately, and obtain electron conductivity $\sigma_{e}$ and hole conductivity $\sigma_{h}$. The mobility is defined by, $\mu_{\alpha\beta}=\sigma_{\alpha \beta}/nq$, where $n$ is the carrier concentration.  The second term in Eq.~\ref{jc} represents the contribution to the electrical current from the temperature gradient and the coefficient $L_{12}$ writes,
\begin{equation}
L_{12}=-\frac{q}{\Omega T N_{n\mathbf{k}}}\sum_{n\mathbf{k}}\mathbf{v}_{n\mathbf{k}\alpha}\mathbf{v}_{n\mathbf{k}\beta}\tau_{n\mathbf{k}}\left( \epsilon_{n\mathbf{k}}-\mu\right)\frac{\partial f_{n\mathbf{k},0}}{\partial \epsilon_{n\mathbf{k}}}.
\label{l12}
\end{equation}
The Seebeck coefficient is defined by the ratio of $L_{12}$ and $L_{11}$,
\begin{equation}
\begin{split}
&S_{\alpha \beta}=\frac{L_{12}}{L_{11}}\\
&=\frac{1}{qT}\frac{\sum_{n\mathbf{k}}\mathbf{v}_{n\mathbf{k}\alpha}\mathbf{v}_{n\mathbf{k}\beta}\tau_{n\mathbf{k}}\left( \epsilon_{n\mathbf{k}}-\mu\right)\frac{\partial f_{n\mathbf{k},0}}{\partial \epsilon_{n\mathbf{k}}}}{\sum_{n\mathbf{k}}\mathbf{v}_{n\mathbf{k}\alpha}\mathbf{v}_{n\mathbf{k}\beta}\tau_{n\mathbf{k}}\frac{\partial f_{n\mathbf{k},0}}{\partial \epsilon_{n\mathbf{k}}}}.
\label{seebeck}
\end{split}
\end{equation}
Note that the Seebeck coefficient is not an additive quantity thus the accumulated Seebeck coefficient is ill-defined. Nevertheless, we can still define a truncated Seebeck coefficient
by changing the condition for the summation both in the numerator and denominator  from \{$n\mathbf{k}$\} to \{$n\mathbf{k},|\mathbf{v}_{n\mathbf{k}}|\tau_{n\mathbf{k}}<\lambda$\}. Effectively, we are able to calculate the contribution to the Seebeck coefficient of electrons with mean free paths up to a given value $\lambda$. The truncated power factor is defined in the same fashion by setting a maximum mean free path for all summations.
The first term in Eq.~\ref{j} corresponds to the heat flow due to the electrochemical potential gradient and the coefficient
$L_{21}=T \,L_{12}$. The second term in Eq.~\ref{j} describes the diffusion of electron under a temperature gradient, where the coefficient $L_{22}$ is defined as,
\begin{equation}
L_{22}=-\frac{1}{\Omega T }\sum_{n\mathbf{k}}\mathbf{v}_{n\mathbf{k}\alpha}\mathbf{v}_{n\mathbf{k}\beta}\tau_{n\mathbf{k}}\left( \epsilon_{n\mathbf{k}}-\mu\right)^2\frac{\partial f_{n\mathbf{k},0}}{\partial \epsilon_{n\mathbf{k}}}.
\label{l22}
\end{equation}
Substitute the electrochemical potential gradient in Eq.~\ref{j} in terms of temperature gradient using Eq.~\ref{jc}, from which we know the electronic thermal conductivity $k_e$ is given by,
\begin{equation}
\kappa_e = L_{22}-L_{21}L_{12}L_{11}^{-1}.
\end{equation}
At high temperatures, the bipolar thermal conductivity can be significant and in current formalism. The bipolar thermal conductivity is written as,
\begin{equation}
\kappa_{bp}=\frac{\sigma_e \sigma_h}{\sigma_e+\sigma_h}\left( S_e-S_h\right)^2 T,
\end{equation}
where we define the electron Seebeck coefficient $S_e$ (hole Seebeck coefficient $S_h$) by only summing over electrons (holes) in the numerator and denominator in Eq.~\ref{seebeck}.
All the above transport properties require the knowledge of electron-phonon scattering and the details to calculate it are as follows.

\subsection{Electron-phonon scattering rate}

The electron-phonon self-energy based on Migdal approximation\cite{migdal1958interaction} is defined by,
\begin{equation}
\begin{split}
&\Sigma_{n\mathbf{k}}=\sum_{m\nu q}\big| g_{mn}^\nu\left(\mathbf{k},\mathbf{q}\right)\big|^2\\
&\left[
\frac{n_{\nu \mathbf{q}}+f_{m\mathbf{k}+q}}{\epsilon_{n\mathbf{k}}-\epsilon_{m\mathbf{k}+\mathbf{q}}+\hbar\omega_{\nu\mathbf{q}}-i\eta} + \frac{n_{\nu \mathbf{q}}+1-f_{m\mathbf{k}+q}}{\epsilon_{n\mathbf{k}}-\epsilon_{m\mathbf{k}+\mathbf{q}}-\hbar\omega_{\nu\mathbf{q}}-i\eta}\right],
\end{split}
\label{selfe}
\end{equation}
where $g_{mn}^\nu\left(\mathbf{k},\mathbf{q}\right)$ is the electron-phonon coupling matrix element and $n_{\nu \mathbf{q}}$ is the phonon distribution. $\epsilon_{n\mathbf{k}}$ is the electron energy and $\omega_{\nu\mathbf{q}}$ is the phonon frequency.
The electron-phonon coupling matrix is given by,
\begin{equation}
g_{mn}^{\nu}\left(\mathbf{k}, \mathbf{q} \right)=\left( \frac{\hbar}{2m_0\omega_{\nu\mathbf{q}}}\right)^{1/2}\left< \psi_{m\mathbf{k}+\mathbf{q}}\Big|\frac{\partial  V_\mathrm{SCF}}{\partial \mathbf{u}_{\nu\mathbf{q}}}\cdot \mathbf{e}_{\nu\mathbf{q}} \Big| \psi_{n\mathbf{k}}\right>,
\label{eph}
\end{equation}
where $m_0$ is the electron rest mass, $\psi_{n\mathbf{k}}$ is the electron wavefunction. $\partial V_\mathrm{SCF}/\partial \mathbf{u}_{\nu\mathbf{q}}\cdot \mathbf{e}_{\nu\mathbf{q}}$ is the first-order variation of the self-consistent potential energy due to the presence of a phonon, as depicted in the density functional perturbation (DFPT) formalism\cite{PhysRevB.76.165108}\cite{RevModPhys.73.515}.
The electron-phonon scattering rate can be calculated from the imaginary part
of self-energy $\Sigma_{n\mathbf{k}}$ by $\Gamma_{n\mathbf{k}}=1/\hbar\,\mathrm{Im}\,\Sigma_{n\mathbf{k}}$. The explicit form of the electron-phonon scattering rate can be written as,
\begin{equation}
\begin{split}
\Gamma_{n\mathbf{k}}&=\sum_{m\nu q}\frac{\pi}{\hbar}\big| g_{mn}^\nu\left(\mathbf{k},\mathbf{q}\right)\big|^2\\
&\times\Big[ \left(n_{\nu \mathbf{q}}+1-f_{m\mathbf{k}+q}\right)\delta\left( \epsilon_{n\mathbf{k}}-\epsilon_{m\mathbf{k}+\mathbf{q}}-\hbar\omega_{\nu\mathbf{q}}\right)\\
&+\left(n_{\nu \mathbf{q}}+f_{m\mathbf{k}+q}\right)\delta\left( \epsilon_{n\mathbf{k}}-\epsilon_{m\mathbf{k}+\mathbf{q}}+\hbar\omega_{\nu \mathbf{q}}\right)\Big].
\end{split}
\label{gamma}
\end{equation}
The inverse of the scattering rate gives the relaxation time,  $\tau_{n\mathbf{k}}=1/\Gamma_{n\mathbf{k}}$.
\subsection{The screening effect of free carriers}

The contribution to the electron-phonon coupling from the long-range polar Fr\" ohlich interaction, which gives rise to the polar optical phonon  (POP) scattering, is expressed by,
\begin{equation}
\begin{split}
g_{n\mathbf{k}, \nu\mathbf{q}}^{m\mathbf{k}+\mathbf{q},\,\mathrm{long}} &= \frac{i e^2}{\Omega \epsilon_0} \sum_{s, \mathbf{G \neq -q}} \left( \frac{\hbar}{2NM_s\omega_{p\mathbf{q}}}\right)^{1/2}\\
&\times \frac{\mathbf{(q+G)} \cdot \mathbf{Z}_s^* \cdot \mathbf{e}_{sp}(\mathbf{q})}{\mathbf{(q+G)}
\cdot {\epsilon(\omega,\mathbf{q})} \cdot \mathbf{(q+G)}}
\left<m\mathbf{k}+\mathbf{q}|e^{i (\mathbf{q+G})\cdot \mathbf{r}}|n\mathbf{k}
\right>,
\end{split}
\end{equation}
where $\mathbf{G}$ is the reciprocal lattice vector, $\mathbf{Z}_s^*$ is the Born effective charge and $\epsilon_0$ is the vacuum permittivity\cite{PhysRevB.95.075206}\cite{PhysRevB.92.054307}\cite{PhysRevLett.115.176401}.
If one performs the Fourier transform of the Poisson's equation, the dielectric constant of a material $\epsilon(\omega,\mathbf{q})$ is interpreted as a function of frequency and wavevector. In the carrier concentration range to our interest ($n \sim \mathrm{10^{19} \,cm^{-3}}$), the plasma frequency estimated using $\omega_p = \sqrt{\frac{ne^2}{\epsilon_\infty \epsilon_0 m}}$ is about 4.9 $\omega_\mathrm{TO}$. That is to say, the free electrons respond to motion of ion so rapidly that we can take the static limit ($\omega \to 0$) of the dielectric constant. The reduced form of dielectric constant $\epsilon(q)$ is known as the Lindhard dielectric function, viz.,
\begin{equation}
\epsilon(q)=1+\frac{1}{2}\frac{k^2_{TF}}{q^2}+\frac{1}{2}\frac{k_{TF}^2}{q^2}\frac{k_F}{q}\left( 1-\frac{q^2}{4k_F^2}\right)\mathrm{ln}\left|\frac{2k_F+q}{2k_F-q} \right|,
\end{equation}
where the Thomas-Fermi screening wavevector is defined by $k_{\mathrm{TF}}^2 =\frac{e^2}{\epsilon_0\epsilon_\infty} \frac {\partial n}{\partial E_F}$ and $k_F$ is the Fermi velocity\cite{mahan2013many}. In fact, we find in PbTe that only small $|\mathbf{q}|$ can lead to strong Fr\" ohlich interaction thus the choice of dielectric constant at large $|\mathbf{q}|$ would not affect the accuracy of the transport calculation. Furthermore, we argue that for those LO phonons that induce POP scattering, the phonon wavevector satisfies $|\mathbf{q}|\ll 2 k_F$ in highly-doped PbTe. In such conditions, the Lindhard dielectric can be further reduced to Thomas-Fermi screening model,
\begin{equation}
\epsilon(q) = \epsilon_\infty\left(1+\frac{k_{TF}^2}{q^2}\right),
\label{tf}
\end{equation}
where $\epsilon_\infty$ is ion-clamped (high frequency) macroscopic dielectric constant from DFPT\cite{PhysRevB.55.10355}\cite{PhysRevB.78.121201}.

The dipole field created by ion motion not only induces POP scattering but results in LO-TO splitting\cite{PhysRevB.56.9524}. When the dielectric constant is modulated by screening, meaning that the capability of the free carriers to screen the dipole field changes, the LO-TO splitting is affected accordingly. The nonanalytical force constant responsible for the LO-TO splitting writes,
\begin{equation}
\begin{split}
{}^{\mathrm{na}}\tilde{\Phi}_{s s'}^{\alpha \beta} (\mathbf{q}) &= \frac{4 \pi e^2}{\Omega\epsilon_0} \frac { (\mathbf{q} \cdot \mathbf{Z}_s^*)_\alpha (\mathbf{q} \cdot \mathbf{Z}_{s'}^*)_\beta }{\mathbf{q} \cdot {\epsilon} \cdot \mathbf{q} }.
\label{eq2}
\end{split}
\end{equation}
In the standard DFPT formalism, the dielectric constant is regarded as a constant, which is valid for a bulk calculation\cite{RevModPhys.73.515}. However, good thermoelectric materials are usually highly doped semiconductors. The electron dynamics is modified profoundly by the free carriers. Therefore, we believe it to be necessary to introduce a carrier-concentration-dependent dielectric constant based on Thomas-Fermi screening model by replacing $\epsilon$ in Eq.~\ref{eq2} with $\epsilon(q)$ defined by Eq.~\ref{tf}.
\subsection{Thermal transport properties}

%%%%%%%%%%%%%%%% mark here %%%%

The heat flux by phonons is caused by the deviation of the distribution function from equilibrium in an isotropic material\cite{PhysRevB.72.014308},
\begin{equation}
\mathbf{J}_{ph}=\frac{1}{\Omega N_{\nu\mathbf{q}}}\sum_{\nu\mathbf{q}}\hbar \omega_{\nu\mathbf{q}}\mathbf{v}_{\nu\mathbf{q}}\left(n_{\nu\mathbf{q}}-n_{\nu\mathbf{q},0}\right).
\end{equation}
Considering the Fourier's law $\mathbf{J}_{ph}=-\kappa_{ph}\nabla_\mathbf{r}T$, we find that the expression for phonon thermal conductivity from linearized Boltzmann transport equation under the relaxation time approximation is,
\begin{equation}
\kappa_{ph}^{\alpha\beta}=\frac{1}{\Omega N_{\nu\mathbf{q}}}\sum_{\nu\mathbf{q}}\frac{\left(\hbar \omega_{\nu\mathbf{q}}\right)^2}{k_B T^2}n_{\nu\mathbf{q}} \left( n_{\nu\mathbf{q}} + 1\right)\mathbf{v}_{\nu\mathbf{q}}^\alpha\mathbf{v}_{\nu\mathbf{q}}^\beta\tau_{\nu\mathbf{q}},
\end{equation}
where $N_\mathbf{q}$ is number of the $q$ point. The calculation of the thermal conductivity requires the phonon dispersion relation, which contains the information of phonon frequency and group velocity. We also need to calculate the relaxation time and this can be calculated by,
\begin{equation}
\frac{1}{\tau_{\nu\mathbf{q}}}=\frac{1}{N_\mathbf{q}}\left( \sum_{\substack{\nu'\nu''\\\mathbf{q}'\mathbf{q}''}}{}^{+}\Gamma^{\nu\nu'\nu''}_{\mathbf{q}\mathbf{q}'\mathbf{q}''}+\frac{1}{2}\sum_{\substack{\nu'\nu''\\\mathbf{q}'\mathbf{q}''}}{}^{-}\Gamma^{\nu\nu'\nu''}_{\mathbf{q}\mathbf{q}'\mathbf{q}''}\right).
\end{equation}
The term ${}^{\pm}\Gamma^{\nu\nu'\nu''}_{\mathbf{q}\mathbf{q}'\mathbf{q}''}$ corresponds to the phonon absorption/emission process,
\begin{equation}
\begin{split}
&{}^{\pm}\Gamma^{\nu\nu'\nu''}_{\mathbf{q}\mathbf{q'}\mathbf{q''}}=\frac{\hbar \pi}{4\omega_{\nu\mathbf{q}}\omega_{\nu'\mathbf{q'}}\omega_{\nu\mathbf{q''}}}\\
&\times\Big| {}^{\pm}V^{\nu\nu'\nu''}_{\mathbf{q}\mathbf{q'}\mathbf{q''}}\Big|^2\begin{bmatrix} n_{\nu'\mathbf{q'}}- n_{\nu''\mathbf{q''}}\\ n_{\nu'\mathbf{q'}}+ n_{\nu''\mathbf{q''}}+1\end{bmatrix}\delta\left(\omega_{\nu\mathbf{q}}\pm \omega_{\nu'\mathbf{q'}}-\omega_{\nu''\mathbf{q''}}\right),
\end{split}
\end{equation}
where ${}^\pm V^{\nu\nu'\nu''}_{\mathbf{q}\mathbf{q'}\mathbf{q''}}$ is the scattering matrix element.

\subsection{Calculation details}
We carried out the first-principles calculation on electronic band structure using a 6 $\times$ 6 $\times$ 6 Monkhorst-Pack $k$-grid with cutoff energy of 70 Ry. We choose the norm-conserving fully relativistic pseudopotentials with local density approximation (LDA)  for exchange-correlation energy functional. The calculation includes the spin-orbit coupling, implemented in Quantum ESPRESSO package\cite{qe}. The lattice constant used in calculation is 6.29 \AA. The band gap given by the DFT calculation is 0.15 eV. We rigidly shift the conduction bands to match the band gap at room temperature, which is 0.316 eV\cite{ravich2013semiconducting}. The doping is modeled with a rigid band model approximation and dopants are assumed to be fully ionized in the whole temperature range in calculation. Given the number of dopants, the chemical potential is obtained solving the charge neutrality equation.

To calculate the dynamical matrix for phonons, we use DFPT with a $6 \times 6 \times 6$ $q$-point mesh. In order to match the bulk phonon dispersion with neutron scattering, a Born effective charge of $Z^*$ = 5.8 and a dielectric constant of 32 from Ref.\cite{PhysRevB.87.115204} are adopted while in the electron phonon coupling calculation, the phonon eigenvalues are directly obtained from DFPT without any modification. To evaluate the electron-phonon coupling matrix, we interpolate the electron-phonon coupling matrix on a $200\times200
\times200$  $k$-point mesh and a $100\times100\times100$ $q$-point mesh using the EPW code\cite{epw}.
To calculate the thermal conductivity, we use a cubic supercell that contains 64 atoms to obtain scattering matrix in the formalism proposed by Ref.\cite{PhysRevB.84.085204}.

We would also like to address the effect of temperature on transport properties. There are several ways to calculate the band structure considering the temperature effect. The most straightforward way is the  \textit{ab initio} molecular dynamics\cite{doi:10.1063/1.4858195}. Some also points out the importance of including the higher-order electron-phonon interaction to capture the correct temperature-dependent band energy\cite{schluter1975pressure}. We adopt the temperature-dependent band gap from experiment\cite{ravich2013semiconducting} and compare the results with constant-band-gap calculation. We realize that the different between the two cases are insignificant\cite{anote}. Consequently, we apply the same lattice constant and band gap for all calculations and the temperature effect is encoded in the distribution functions of electrons and phonons.
%Here, we focus on the scattering process for electrons rather than the exact description of the band structure. For simplicity,  The effective mass and band gap are thus treated as temperature-independent.

\section{Results and discussion}
\subsection{The effect of screening}
\begin{figure}[t]
\includegraphics[width=\linewidth]{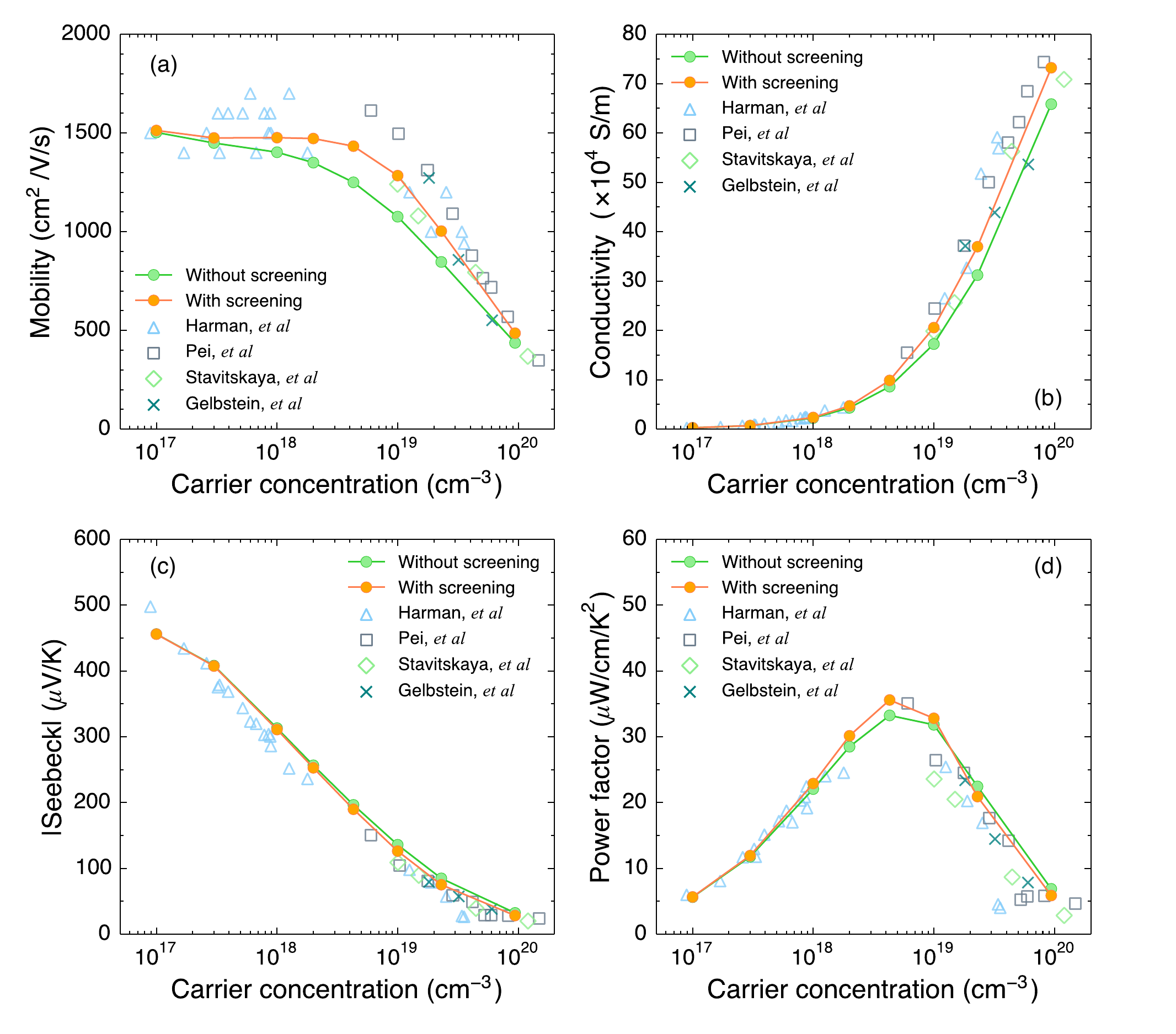}
 \caption{(a) The mobility, (b) the electrical conductivity, (c) the Seebeck coefficient, and
(d) the power factor of PbTe as a function of carrier concentration at 300 K with and without considering the screening effect. Dotted lines are simulation and isolated dots are experimental value. The triangles are from Ref.\cite{harman1996high}, squares from Ref.\cite{AENM}, diamonds from Ref.\cite{stavitskaya1966ba}, and crosses from Ref.\cite{Gelbstein2001}.}
\label{fig2}
\end{figure}
The electron transport properties for $n$-type PbTe at room temperature with/without considering the screening effect of the free carriers are demonstrated in Fig.~\ref{fig2}. The free carriers in doped PbTe screen the dipole field generated by ion vibration outside the sphere defined by the the screening length. One thus expects reduced POP scattering and higher mobility after taking into account the screening effect. We notice from Fig.~\ref{fig2} that considering the screening effect does yield higher mobility and conductivity than without the screening effect, as well as a more desirable agreement with experiment even though the discrepancies in Seebeck coefficient and power factor for the two cases are not significant.

\begin{figure}[b]
\includegraphics[width=\linewidth]{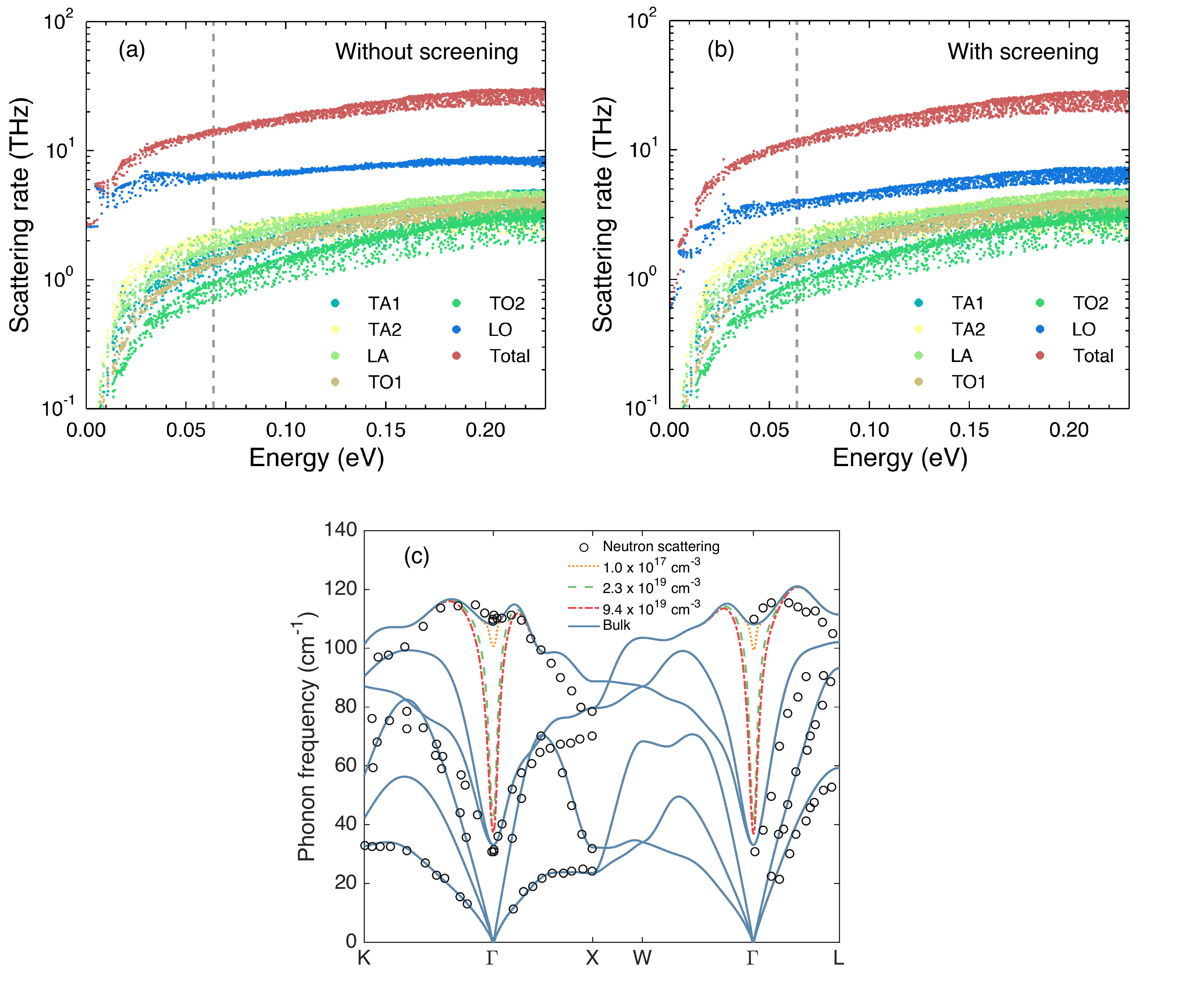}
\caption{ (a-b) The energy-resolved electron-phonon scattering rates for conduction band electrons due to phonon modes of different branches at 300 K with/without considering the screening effect. The zero energy marks the conduction band minimum and the dashed line indicates the location of chemical potential. The dopant concentration is 2.3 $\times 10^{19}\, \mathrm{cm}^{-3}$. (c) The phonon dispersion for different free carrier concentration compared with neutron scattering experiment\cite{Cochran433}.}
\label{fig1}
\end{figure}
\begin{figure}[t]
\includegraphics[width=\linewidth]{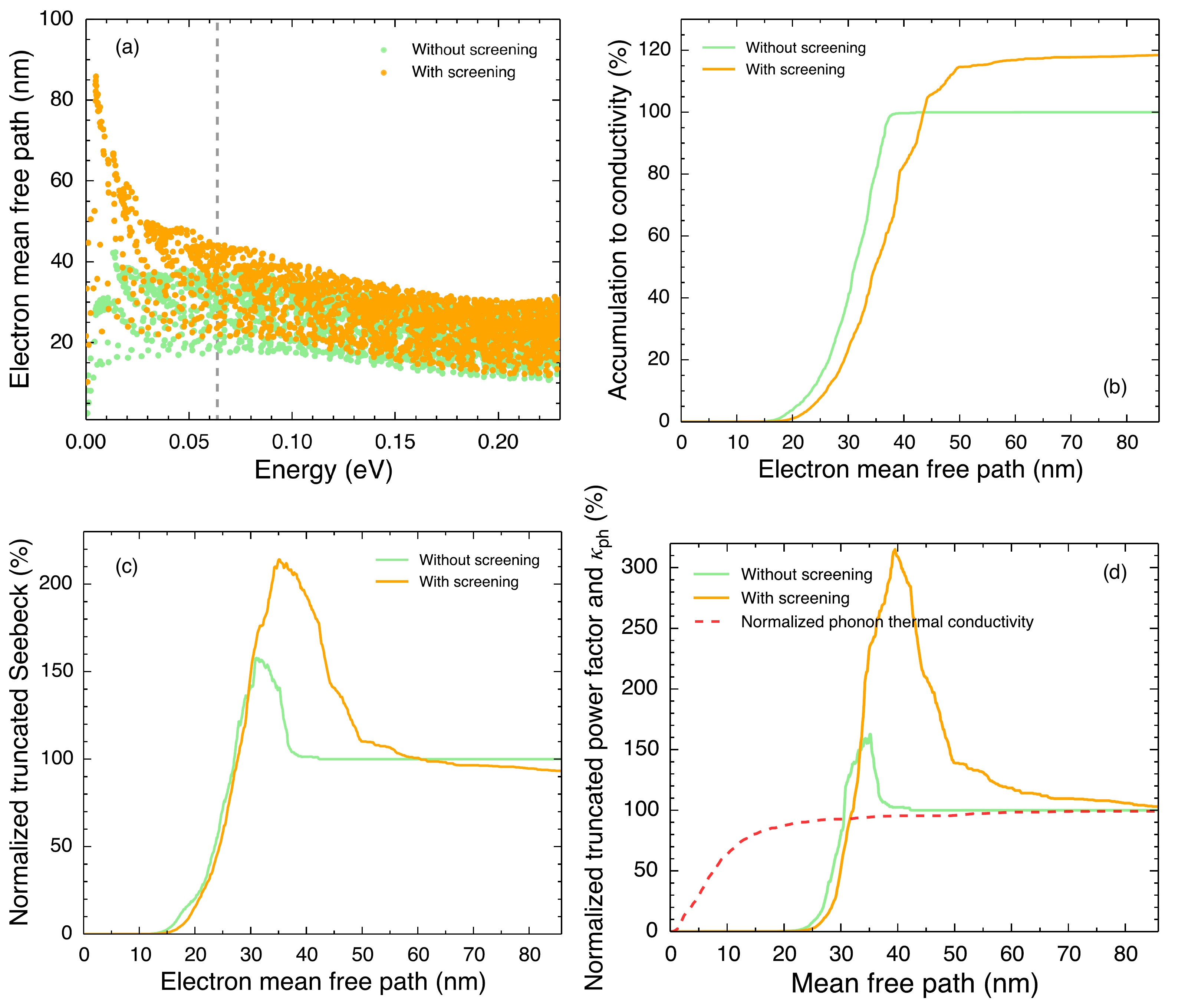}
 \caption{(a) The electron mean free path as a function of energy with and without considering the screening effect. The dashed line indicates the chemical potential and zero energy indicates the conduction band minimum. (b) The accumulated electrical conductivity with respect to electron mean free path. (c) The normalized truncated  Seebeck coefficient with respect to the electron mean free path.   (d) The normalized truncated power factor with respect to electron mean free path compared with normalized accumulated phonon thermal conductivity with respect to phonon mean free path. The dopant concentration is 2.3 $\times 10^{19}\, \mathrm{cm}^{-3}$.}
\label{wwos}
\end{figure}
The electron-phonon scattering rates due to different phonon branches without considering the screening effect are shown in Fig.~\ref{fig1} (a). The scattering due to LO phonons is stronger than other phonon branches. Near the conduction band minimum, only phonon absorption process is allowed to happen. When electron energy is larger than LO phonon energy, both phonon absorption and emission can happen. This leads to the sudden jump of the scattering rate near 0.01 eV. Note that the dielectric constant calculated from DFPT is 104, which is overestimated compared with experiment reported value 32\cite{Cochran433}. As a result, the LO phonon frequency is 10 meV from calculation, which is underestimated than value of the value of 13 meV found through neutron scattering experiment\cite{Cochran433}. One interesting feature of the scattering rate due to LO phonon is the relatively weak energy dependence. The scattering due to LO phonon consists of two
contributions: the non-polar optical phonon deformation potential (ODP) scattering and POP scattering, as discussed in detail in Ref.\cite{lundstrom2009fundamentals}. The non-polar ODP scattering rate scales with $\sqrt{E}$ ($E$ is the electron energy measured from conduction band edge), and the POP scattering rate scales with $\mathrm{sinh}^{-1}(\sqrt{E})$ assuming a parabolic band. The actual non-parabolic band structure of PbTe might change the exact energy dependence of scattering rates. Still qualitatively, the POP scattering rate increases less rapidly with increasing electron energy than the non-polar ODP scattering rate.

When including the screening effect shown in Fig.~\ref{fig1} (b), the scattering rate due to LO phonon decreases especially for low-energy electrons ($E <  0.1\, \text{eV}$). At the conduction band minimum, because of the screening effect, the scattering rate due to LO phonon decreases from 2.5 THz to 0.5 THz. On the contrary, the reduction is much less noticeable for high-energy electrons. The screening effect in principal should only be able to affect POP scattering rather than ODP scattering. For high-energy electrons ($E >  0.1\, \text{eV}$), non-polar ODP scattering is much stronger than POP scattering so that the reduction in POP scattering becomes less discernible than low-energy electrons.

Another consequence resulting from the screening effect is the weakened LO-TO splitting. In Fig.~\ref{fig1} (c), we clearly observe that as the carrier concentration increases, the gap between LO and TO phonons near zone center is progressively narrowed. In the high-carrier-concentration limit and the long-wavelength limit, one should no longer be able to distinguish a LO and TO phonon since the screening length has becomes so small that the long-range dipole field responsible for the LO-TO splitting vanishes. The convergence of long-wavelength LO and TO phonon reminds us to examine whether it gives rise to stronger anharmonicity since TO phonon contributes remarkably to phonon-phonon scattering in PbTe\cite{Delaire2011}. However, we do not observe any noticeable difference after carrying out thermal conductivity calculation, because only small fraction of LO phonons become TO phonons such that the three-phonon scattering phase space is barely modified.

Fig.~\ref{wwos} (a) shows that the mean free path for low-energy electrons increases dramatically when including the screening effect. Although the LO phonon scattering is the prominent scattering source (both for cases with/without screening), it is not strong enough to totally overshadow the contribution from TO phonons and acoustic phonons. For electrons with energy near the chemical potential, the scattering of non-LO phonons contributes to the total scattering comparably with LO phonons. Note the electron group velocity near the chemical potential is weakly dependent on energy, as also shown in Ref.\cite{Chen2013}. That is to say, the electron mean free path is a monotonically decreasing function of energy near the chemical potential because the electron relaxation time decreases monotonically with increase in energy.

The electrical properties of electron as a function of electron mean free path is displayed in Fig.~\ref{wwos} (b), where the conductivity is enhanced by about 20 \% due to the screening effect. Besides, the mean free path spectrum is shifted to higher values. An interesting feature is found in the truncated Seebeck coefficient in Fig.~\ref{wwos} (c): up to certain mean free path, the truncated Seebeck coefficient can be even higher than the total Seebeck coefficient. As is known, above the chemical potential, the electrons contribute dominantly to the Seebeck coefficient with negative signs. However, the electrons below the chemical potential have positive signs and they cancel the contribution from electrons above the chemical potential. Recall the mean free path is almost a monotonically decreasing function of energy in Fig.~\ref{wwos} (a). The above observation then translates to the fact that that the long-mean-free-path electrons contributes contribute  ``negatively'' to Seebeck whilst the short-mean-free-path ones contribute ``positively'', which explains the emergence of the peak in the truncated Seebeck coefficient at a critical mean free path. The screening effect pushes the critical mean free path from 35 nm to 40 nm and the peak value rises from 160 \% to 210 \%. In Fig.~\ref{wwos} (d) the peak power factor at the critical mean free path is as high as 310 \% with screening effect and 160 \% without screening effect. We also find that the truncated power factor with/without screening effect at the long-mean-free-path limit are almost the same, albeit the screening effect greatly alters the mean free path distribution.
\begin{figure}[b]
\includegraphics[width=0.85\linewidth]{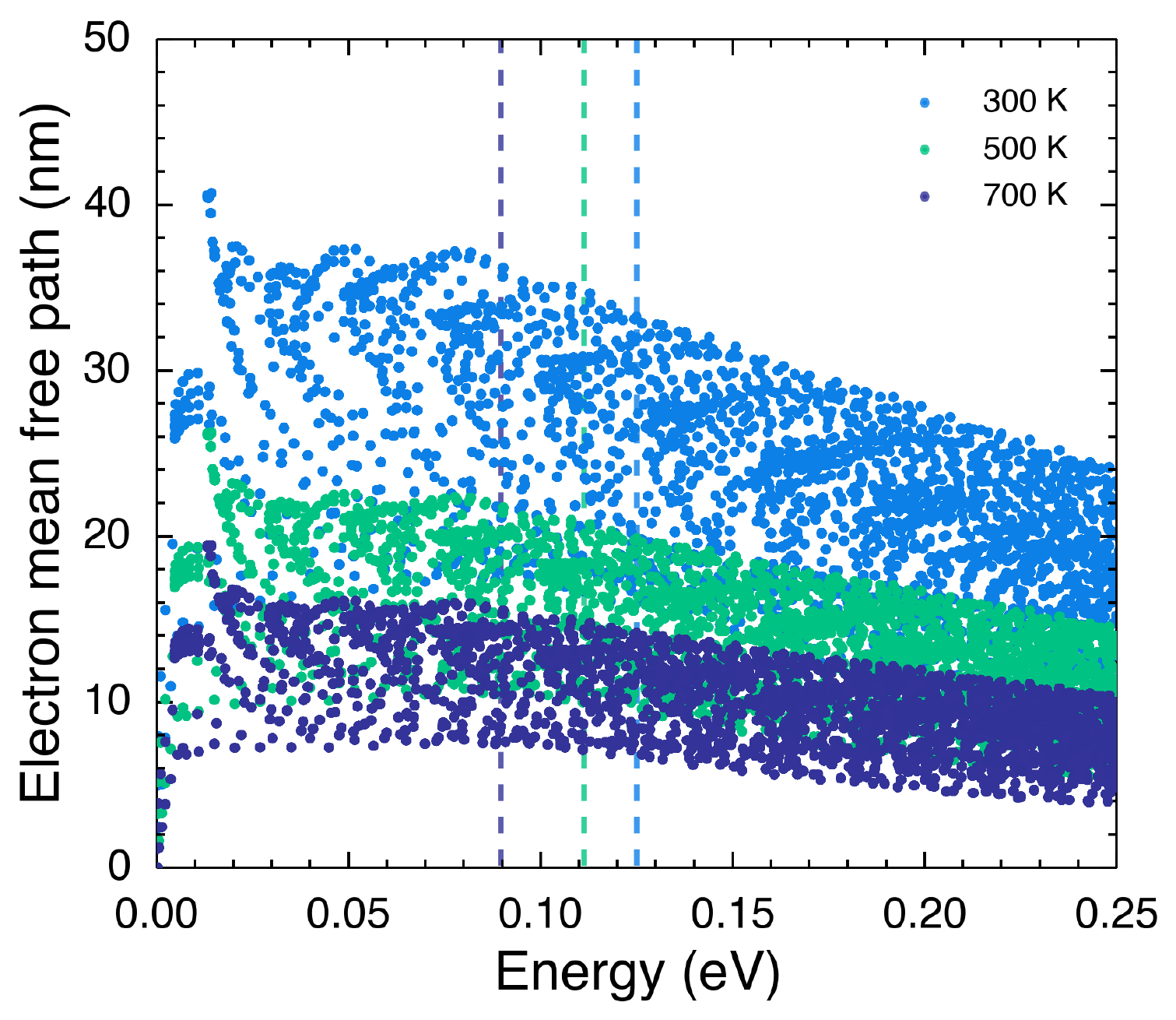}
\caption{The electron mean free path as a function of energy at different temperatures. The dashed line indicates the chemical potential and zero energy corresponds the conduction band minimum. The dopant concentration is 5.8 $\times 10^{19}\, \mathrm{cm}^{-3}$. }
\label{fig22}
\end{figure}
\begin{figure}
\includegraphics[width=\linewidth]{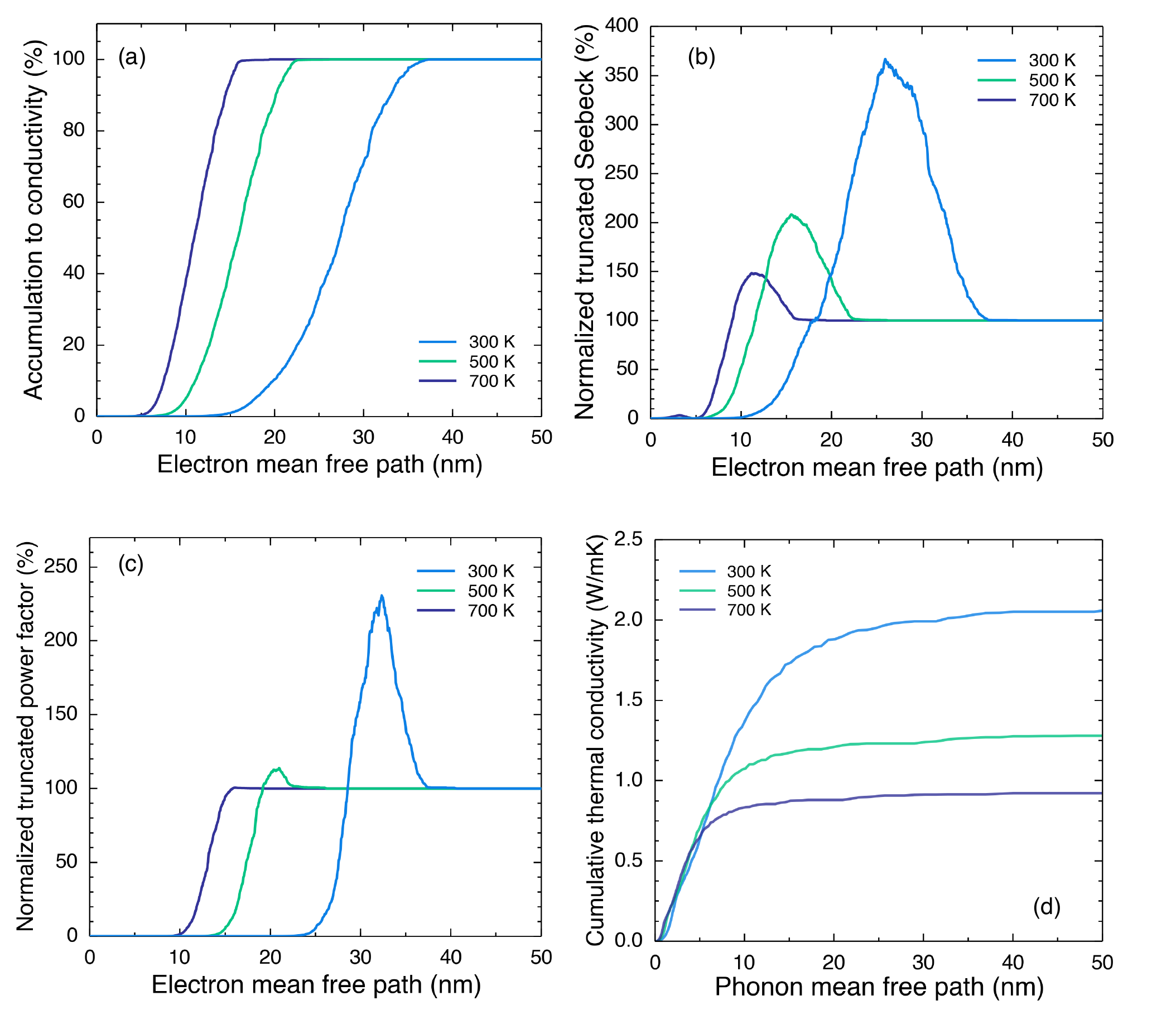}
\caption{(a) The accumulated electrical conductivity with respect to electron mean free path. The normalized truncated (b) Seebeck coefficient and (c) power factor with respect to electron mean free path. (d) The accumulated  lattice thermal conductivity with respect to phonon mean free path. The dopant concentration is 5.8 $\times 10^{19}\, \mathrm{cm}^{-3}$. }
\label{fig3}
\end{figure}
\begin{figure*}
\includegraphics[width=0.85\linewidth]{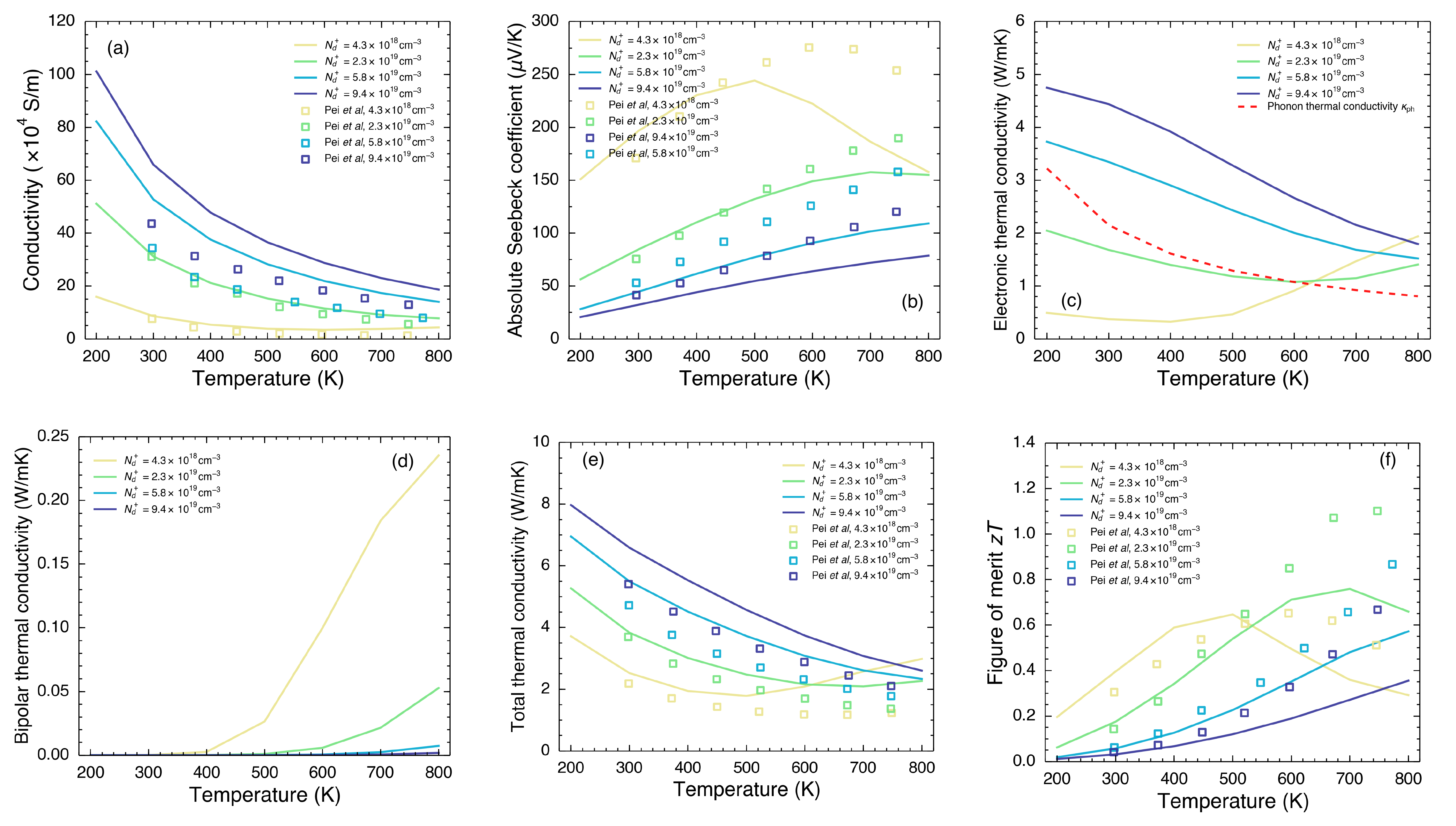}
 \caption{(a) The conductivity, (b) the Seebeck coefficient, (c) the electronic thermal conductivity compared with phonon thermal conductivity, (d) the bipolar thermal conductivity, (e) the total thermal conductivity and (f) the figure of merit $zT$ of PbTe as a function of temperature for different ionized donor
concentrations. The squares are experimental results from Ref.\cite{AENM}}
\label{fig5}
\end{figure*}

If we compare the truncated power factor to the cumulative thermal conductivity, we first realize that the major contribution to thermal conductivity is from phonons with mean free paths smaller than 20 nm, whilst for the truncated power factor, most contributions are from electrons with mean free paths higher than 20 nm. Surprisingly, this finding contradicts with the general case where the electron means free paths are much smaller than the phonon mean free paths, which emphasizes the importance of considering both electron and phonon when designing the nanostructures for PbTe. However, despite that nanostructures that scatter phonons may also scatter electrons, the long-mean-free-path electrons contributed negatively to the Seebeck coefficient. Nanostructures may scatter these long-mean-free-path electrons, leading to actually increased Seebeck coefficient and decreased electrical conductivity.  This is seen in some of past experiments\cite{tan2016non}, although arguably, we cannot tell at this stage if these past experimental observations is due to filtering of long-mean-free-path electrons or the thermionic effect\cite{PhysRevLett.92.106103}.

\subsection{The effect of temperature}
We proceed to study the temperature dependence of the transport properties. When raising the temperature, the mean free paths not only decreases but covers a narrower mean free path range, shown in Fig.~\ref{fig22}. For electrons with the same energy, the mean free path is not a single value but forming a ``band'' containing a series of possible values. The width of the ``band'' shrinks with rising temperature. At elevated temperatures, the population of phonons scales with $T$. From the analysis of our calculation, the scattering rate is directly related to temperature in a power law $\propto T$. Hence, for the electrons with the same energy, the scattering rates are rescaled by the temperature, and the inverse of the scattering --- the relaxation time, will decrease and spread in a narrower region, causing a narrower ``band''. In addition, at higher temperatures, the chemical potential shifts towards the band minimum. This is because the Fermi-Dirac distribution function spreads wider in the energy scale as temperature rises. To match the fixed amount of positively charged ionized donors, the chemical potential must be lowered.

The electrical properties of electron as a function of electron mean free path at different temperatures are displayed in Fig.~\ref{fig3} (a). At room temperature, the greatest contribution to the conductivity comes from the electrons with mean free paths smaller than 37 nm, regarded as the maximum electron mean free path. As temperature is lifted, the maximum mean free path decreases. We also realize that the height of the peak in normalized truncated Seebeck coefficient decreases when temperature rises, as described in Fig.~\ref{fig3} (b). We refer to electron mean free path at different temperatures in Fig.~\ref{fig22}. At room temperature, the mean free paths of electrons above and below the chemical potential contrast profoundly. At high temperatures, such contrast gradually becomes inconspicuous, causing lowered peak height in the normalized truncated Seebeck coefficient. In Fig.\ref{fig3} (c), the maximum normalized truncated power factor, is as high as 230 \% at 300 K but the maximum is almost unity at 700 K. Compared with phonon mean free path profile in Fig.\ref{fig3} (d), the mean free paths of electrons that contribute majority of the power factor are larger than phonons contributing majority of the thermal conductivity at all temperatures.

The Fig.~\ref{fig5} (a) presents the conductivity as a function of temperature for different dopant concentrations, compared with La-doped PbTe from the experiment\cite{AENM}. The decrease of the mobility versus temperature is mostly because electron-phonon scattering becomes stronger with increase in temperature. Our calculation overall captures the correct trend both for mobility and conductivity. However, the calculated Seebeck coefficient in Fig.~\ref{fig5} (b) is largely underestimated above 400 K for the lowest dopant concentration. As is known in PbTe, high temperatures flatten the band structure near the band edge, causing a larger effective mass\cite{Ravich1971a}, while in calculation, the band structure keeps unchanged. This also leads to the discrepancy between the calculation of the calculated power factor and experimental results. The band gap and the alignment of different valleys are function of temperature in reality that can also alter the Seebeck coefficient, yet not captured by our constant-band-gap calculation. In addition, the experiment demonstrates that for La-doped PbTe when the Hall carrier concentration is above $6\times10^{19} \mathrm{cm}^{-3}$, there is deviation from valence counting rule that each dopant atom provides one carrier\cite{AENM}. We believe this further contributes the differences between our calculation and experiment at $9.4\times 10^{19} \mathrm{cm}^{-3}$.

Since our first-principles calculation of electron-phonon scattering is parameter-free, we can calculate the electronic thermal conductivity at different temperatures for different dopant concentrations instead of relying on the Wiedemann-Franz law, shown in Fig.~\ref{fig5} (c). For high dopant concentrations, even though the chemical potential is being lowered towards the band minimum as the temperature is elevated, the chemical potential is still close to the conduction band. For the low carrier concentration case ($4.3 \times 10^{-18}\,\mathrm{cm}^{-3}$), the chemical potential is closer to the middle of the band gap. With increasing temperature, holes start to contribute to the electronic thermal conductivity since the bipolar transport becomes noticeable, which corresponds to the increase above 400 K. Note that at the high carrier concentration ($5.8 \times 10^{-19}\,\mathrm{cm}^{-3}$), the thermal conductivity is lower than the electronic thermal conductivity which marks the significance to accurately estimate the electronic thermal conductivity.

In experiment, it's usually difficult to distinguish the bipolar thermal conductivity from the measured thermal conductivity. However, the bipolar thermal conductivity can be explicitly calculated from our DFT calculation, shown in Fig.~\ref{fig5} (d). A noticeable increase is only observed in low concentration of $4.3 \times 10^{-18}\,\mathrm{cm}^{-3}$ above 400 K. The total thermal conductivity is shown in  Fig.~\ref{fig5} (e). Our results in Fig.~\ref{fig5} (c) indicates that the increase of electronic thermal conductivity leads to an increase of the total thermal conductivity. However, the experiment only shows a minor increase. We believe this is due to fact that the calculation does not capture the increased effective mass and temperature-dependent band gap above 400 K.
The figure of merit at high temperature is largely underestimated demonstrated in Fig.~\ref{fig5} (f), again, due to the inaccurate band structure at high temperatures. For the highest dopant concentration, both calculation and experiment show a monotonic increase with temperature because the the chemical potential is still far from being at the middle of the band gap so that the bipolar effect is insignificant.

 \section{Conclusion}

We study the electron-phonon interaction in \textit{n}-type PbTe from first-principles calculation and obtain the electron-phonon scattering rates and electron mean free paths at different temperatures. The LO phonon in PbTe plays an important role in determining the lifetime of electrons. The electron mean free path as a function of energy follows almost the same trend as the relaxation time because of the weak energy dependence of group velocity. This makes the electron mean free path decrease monotonically with energy. The screening effect at high carrier concentrations weakens the LO-TO splitting for phonons and reduces the POP scattering especially for low-energy electron. It also shifts the mean free path distribution towards higher values whilst the integrated transport properties are slightly changed. At elevated temperatures, the scattering rates scale with $T$ and the electron mean free path distribution is shifted towards lower values.

The truncated Seebeck coefficient and power factor as a function of electron mean free path is not a monotonically increasing function. There exists a critical mean free path, corresponding to that of the electrons at the chemical potential, below which electrons contribute positively to the Seebeck coefficient while longer-mean-free-path electrons contribute negatively to the Seebeck coefficient. More interestingly, unlike in silicon, the electron mean free paths in PbTe are not significantly smaller than the mean free paths of most of the phonons. This inspires us to further investigate the scattering by interfaces both for electrons and phonons to rigorously and comprehensively answer the question if nanostructuring works better in silicon than in PbTe.

%means that comprehensive information on the mean free paths of electrons and phonons enables researchers to have a better understanding of the microscopic transport picture both for electrons and phonons. The level of details given by our calculation would benefit the rational design of nanostructures to improve the thermoelectric performances.

\begin{acknowledgments}
This research is supported as part of the Solid-State Solar-Thermal Energy Conversion Center (S3TEC)  an Energy Frontier Research Center funded by the U.S. Department of Energy (DOE), Office of Science, Basic Energy Sciences (BES), under Award No. DE-SC0001299 / DE-FG02-09ER46577 (for fundamental research on electron-phonon interaction in thermoelectric materials), by
the DARPA MATRIX program, under Grant HR0011-16-2-0041 (for developing and applying the simulation codes to support MATRIX team members). The authors thank Q. Xu, Y. Tsurimaki, G. D. Mahan for helpful discussion.

%National Science Foundation (NSF) under Award # <insert grant award number> (insert  research area 2), and by the National Institutes of Health (NIH) under Award # <insert grant award number> (insert research area 3). In addition, Investigator #1 acknowledges support by DOE BES grant # <insert grant award number> (insert research area 4); Investigator #2 acknowledges support by NIH grant # <insert grant award number> (insert research area 5), and the NSF DMR-grant # <insert grant award number> (insert research area 6).?

\end{acknowledgments}

\bibliography{pbte}

\begin{thebibliography}{51}
\bibitem [{ano()}]{anote}%
  \BibitemOpen
  \href@noop {} {\bibinfo  {journal} {We adopt rigid band approximation and
  shift the whole conduction bands relative to the valence bands to match the
  experimental band gap at different temperature. Note the band gap of lead
  telluride turns from direct into indirect above 400 K, which is missed in our
  treatment. In fact, we make sure the band gap in the calculation consistent
  with experimental values although the valley alignment might be inaccurate.
  And we find such treatment gives similar transport properties as a function
  of temperature, compared with the constant-band-gap calculation in the main
  text}\ }\BibitemShut {NoStop}%
\end{thebibliography}
\bibliographystyle{apsrev4-1}

\clearpage
\appendix

\end{document}